\documentclass[lettersize,conference]{IEEEtran}
\usepackage{amsmath,amsfonts}
\usepackage{algorithmic}
\usepackage{algorithm}
\usepackage{array}
\usepackage[caption=false,font=normalsize,labelfont=sf,textfont=sf]{subfig}
\usepackage{textcomp}
\usepackage{stfloats}
\usepackage{url}
\usepackage{verbatim}
\usepackage{graphicx}
\usepackage{cite}
\usepackage[dvipsnames]{xcolor}
\DeclareRobustCommand*{\IEEEauthorrefmark}[1]{\raisebox{0pt}[0pt][0pt]{\textsuperscript{\footnotesize\ensuremath{\ifcase#1\or *\or \dagger\or \ddagger\or%
    \mathsection\or \mathparagraph\or \|\or **\or \dagger\dagger%
    \or \ddagger\ddagger \else\textsuperscript{\expandafter\romannumeral#1}\fi}}}}

\begin{document}

\title{Noise Modeling and Calibration of a Two-Stage Cryogenic Charge Amplifier for the SPLENDOR Experiment}

\newcommand{\nk}[1]{\textcolor{red}{(NK: #1)}}
\newcommand{\ja}[1]{\textcolor{teal}{(JA: #1)}}

\author{
    \IEEEauthorblockN{Jadyn Anczarski\IEEEauthorrefmark{1}\IEEEauthorrefmark{2},
    Owen Andrews\IEEEauthorrefmark{3},
    Taylor Aralis\IEEEauthorrefmark{1},
    Caleb Fink\IEEEauthorrefmark{4}\IEEEauthorrefmark{5}\IEEEauthorrefmark{6},
    Noah Kurinsky\IEEEauthorrefmark{1},\\
    Arran Phipps\IEEEauthorrefmark{3},
    Aditi Pradeep\IEEEauthorrefmark{1},
    Betty A. Young\IEEEauthorrefmark{7}
    }
    \IEEEauthorblockA{\IEEEauthorrefmark{1} Kavli Institute for Particle Astrophysics and Cosmology,
    SLAC National Laboratory,
    Menlo Park, CA 94025, USA}
    \IEEEauthorblockA{\IEEEauthorrefmark{2} Department of Physics,
    Stanford University,
    Stanford, CA 94305, USA}
    \IEEEauthorblockA{\IEEEauthorrefmark{3} Department of Physics,
    California State University - East Bay,
    Hayward, CA , USA}
    \IEEEauthorblockA{\IEEEauthorrefmark{4} Institute for Quantum and Information Science,
    Syracuse University
    Syracuse, NY , USA}
    \IEEEauthorblockA{\IEEEauthorrefmark{5} Department of Physics,
    Syracuse University
    Syracuse, NY , USA}
    \IEEEauthorblockA{\IEEEauthorrefmark{6} Los Alamos National Laboratory,
    Los Alamos, NM , USA}
    \IEEEauthorblockA{\IEEEauthorrefmark{7} Department of Physics,
    Santa Clara University
    Santa Clara, CA 95053 USA}
}

\maketitle

\begin{abstract}
The SPLENDOR Collaboration studies novel narrow-gap semiconductors and engineered a substrate agnostic detector platform to achieve $\mathcal{O}$(meV) energy sensitivity designed for low mass dark matter searches. This was achieved using low-capacitance and low-noise commercial CryoHEMTs in a split-stage topology integrated throughout a dilution refrigerator. Designed with a source-follow HEMT at the base temperature stage and a voltage amplifier at 4\,K, this amplifier has input-limited voltage noise of 10 $\text{nV}/\sqrt{\text{Hz}}$ and current noise of \textbf{100} $\text{aA}/\sqrt{\text{Hz}}$ at 1kHz. In agreement with this noise level and a photon calibration, this amplifier has a $\text{19} \pm \text{4} $ electron resolution. 
\end{abstract}


\section{Introduction}

\IEEEPARstart{P}{robing} the low-mass dark matter regime (keV–MeV masses) requires sensitivity to $\mathcal{O}(\text{meV})$ energy deposits, due in part to the exceedingly low kinetic energy of the dark matter~\cite{essig2023snowmass2021}. Narrow band-gap semiconductors make this possible by enabling electrons to be excited into the conduction band with sub-eV thresholds, allowing tests of electron-recoil dark matter theories and, when combined with other observables such as phonons, improved background discrimination~\cite{CDMS2013}. In directional semiconductors, charge-only detection further enables daily modulation searches that can distinguish a dark matter halo signal from backgrounds~\cite{SPLENDOR}.

Low-noise cryogenic HEMTs (CryoHEMTs) have been widely deployed in particle searches, including CDMS and Ricochet, achieving electron-equivalent resolutions of 91\,$\text{eV}\text{ee}$~\cite{PHIPPS2019181} and 30\,$\text{eV}\text{ee}$~\cite{augier2023demonstration}. However, in these systems the input stages operate at higher temperatures than the detector payload, limiting charge resolution leading to significant design challenges. The SPLENDOR collaboration is pursuing single-electron sensitivity by moving amplification to the base stage of a dilution refrigerator and by developing a substrate-agnostic, split-stage HEMT-based charge amplifier with $\sim 20\,e^-$ resolution~\cite{SPLENDOR}. This design minimizes parasitic capacitance and allows coupling to a wide range of semiconductors without requiring complex target-specific fabrication ~\cite{anczarski2023twostage}.

This letter builds on our initial report~\cite{anczarski2023twostage} by demonstrating true charge-amplifier behavior through the integration of high-resistance cryogenic bias resistors and operation with a silicon detector. Section II describes the full experimental setup, including amplifier, filtering, room-temperature electronics, calibration, and data acquisition. Section III presents amplifier characterization and noise performance. Section IV describes the LED shot-noise calibration, achieving a resolution of $19 \pm 4\,e^-$. Section V compares input-noise predictions to LED-based calibration and discusses limitations, upgrades, and broader applications beyond SPLENDOR’s light dark matter program.

\section{Experimental Setup}

\begin{figure}
    \centering
    \includegraphics[width = 0.53\textwidth]{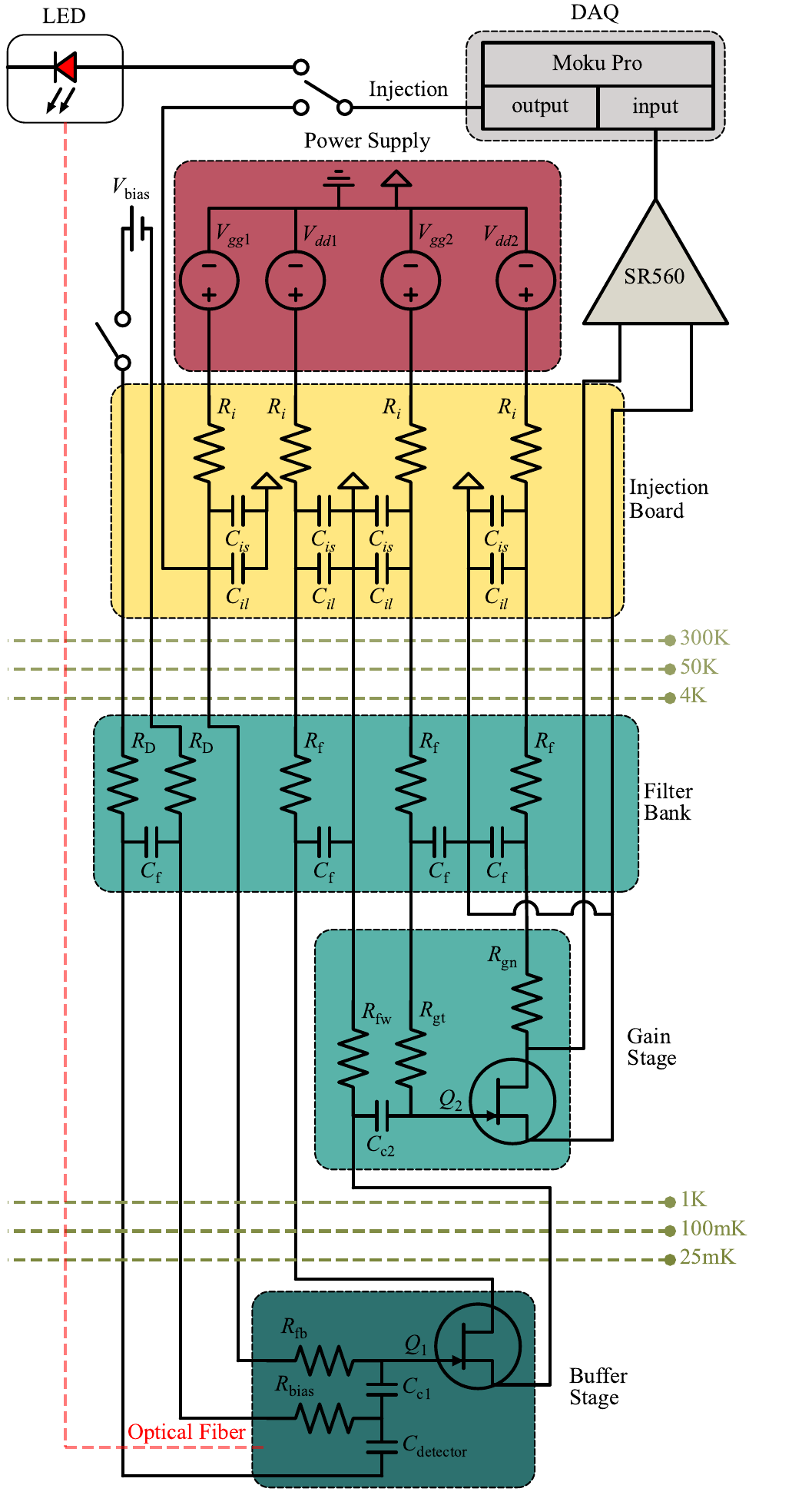}
    \caption{Schematic of the full experimental setup for characterization and pulse detection. Cryogenic stages are shown in dark teal (25\,mK) and light teal (4\,K); at 4\,K the voltage-gain stage includes a filter bank. At room temperature, an injection board (yellow) provides selectable filtering to signal ground or fridge ground. Signal and fridge grounds are connected at the chassis of the power supply. The output signal is amplified by a Stanford Research Systems SR560 to overcome the MokuPro’s noise floor. The MokuPro is a multi-instrument DAQ with input/output control; its output is used either to inject a signal at Q1’s gate for gain and frequency-response measurements or to drive LED pulses which inject light down optical fibers connecting to the base temperature housing. The LED pulses are used for shot-noise calibration. A battery can be connected or disconnected to bias the detector at base temperature.}
    \label{fig: Schematic Offical}
\end{figure}

The SPLENDOR cryogenic split-stage amplifier is shown schematically in Fig.~\ref{fig: Schematic Offical}, with component values summarized in Table~\ref{tab:schematic value}. The design consists of a base-temperature buffer stage and a 4\,K gain stage, both using commercial CryoHEMT devices~\cite{cryoHEMT}.

At base temperature ($\sim$25\,mK), Q1 operates as a transimpedance buffer in a source-follower configuration, integrating charge generated in the detector while isolating the signal from parasitic capacitance of the long cabling up the dilution refrigerator. A low-capacitance ($C_{gs}=1.6$\,pF) HEMT minimizes amplifier input capacitance. The 4\,K stage (Q2) is configured in common-source mode to provide additional gain and incorporates cryogenic filtering. To achieve high transconductance, Q2 uses a higher-capacitance ($C_{gs}=200$\,pF) HEMT, making the buffer stage essential for preserving charge resolution.

The amplifier was tested in SLAC’s Oxford Proteox dilution refrigerator (OLAF)\footnote{A second version of the amplifier was run in a simialr Proteox system at LANL, with similar performance.}. Both HEMTs were biased (Table \ref{tab:hemt_bias}) using a custom low-noise supply we refer to as the HEMT Power Supply (HPS). The base stage dissipated 2.3\,$\mu$W, raising the temperature to 25\,mK from the nominal 10\,mK\footnote{These data were taken during a time period where the cooling power of the DR was reduced. In past runs, heating has driven the DR to temperatures in the 15-20 mK range, consistent with the expected bias power quoted here.}. Heat load from the 4\,K stage (Q2, $<$20\,$\mu$W) was negligible compared to the cooling power at 4\,K. A superconducting shielded NbTi $\mu$D-25 cable connected the 25\,mK buffer to the 4\,K stage, and a copper $\mu$D-25 cable linked the 4\,K stage to the vacuum feedthrough.

At room temperature, a custom four-layer injection board filtered all bias and signal lines, with selective routing to signal and fridge grounds. The HPS provided biasing and defined a star ground. The differential output was amplified with a Stanford Research Systems SR560 (gain $\times 100$) before digitization. Data acquisition was performed with a MokuPro system (Liquid Instruments), which provides 600\,MHz bandwidth and 30\,nV/$\sqrt{\text{Hz}}$ input noise at 100\,Hz. Analysis was conducted using the Python package \texttt{splendaq}~\cite{watkins2023splendaq}.

The amplifier housing supports two independent channels via 12 cryogenic twisted pairs, allowing simultaneous operation with different detector substrates. For the tests reported here, Si chips were mounted in the base-temperature housing (see Fig.~2 of~\cite{anczarski2023twostage}), providing a well-understood capacitance for noise characterization and enabling $625\,$nm LED calibration. This configuration ensures that measured noise includes realistic detector capacitance while permitting absolute calibration of the charge response.

\begin{table}
\caption{Schematic Values For Figure \ref{fig: Schematic Offical}}
\label{tab:schematic value}
\centering
\begin{tabular}{|c|c|c|}
\hline
Variable & Nominal Value & Info \\
\hline
\hline
$V_\text{bias}$ & $9 \text{V}$  & 9V Battery\\

$V_\text{gg1}$ & $0 \text{mV}$  & HEMT Power Supply\\

$V_\text{dd1}$ & $100\text{mV}$ &  HEMT Power Supply\\

$V_\text{gg2}$ & $-130\text{mV}$ & HEMT Power Supply\\

$V_\text{dd2}$ & $1200\text{mV}$ & HEMT Power Supply\\
\hline
\hline
$R_\text{i}$ & $100\Omega$ & Resistor Type\\

 $C_\text{is}$ & $10\text{nF}$ & Ceramic BX\\

 $C_\text{il}$ & $470\mu\text{F}$ & Ceramic X5R\\
\hline
\hline
 $R_\text{D}$ & $1\text{k}\Omega$ & Thin Film\\

 $R_\text{f}$ & $82\Omega$ & Thin Film \\

 $C_\text{f}$ & $22\mu\text{F}$  & Ceramic X5R \\

 $R_\text{fw}$ & $1\text{K}\Omega$ & Thin Film\\

 $R_\text{gt}$ & $100\text{k}\Omega$ & Thin Film\\

 $R_\text{gn}$ & $1\text{k}\Omega$ & Thin Film\\

 $C_\text{c2}$ & $22\mu\text{F}$ & Ceramic X5R\\

 $Q_\text{2}$ & $200\text{pF}$ & CryoHEMT\\
\hline
\hline
 $R_\text{fb}$ & $2\times100\text{M}\Omega$ & MiniSystems\\

  $R_\text{bias}$ & $2\times100\text{M}\Omega$ & Vishay CS44\\

 $C_\text{c1}$ & $100\text{pF}$ & Film Acrylic\\

 $C_\text{det}$ & $\sim 2\text{pF}$ & 500$\mu$m Si Au point contact \\

 $Q_\text{1}$ & $1.6\text{pF}$ & CryoHEMT\\
\hline

\end{tabular}
\end{table}

\begin{table}
\caption{CryoHEMT bias points and dissipated power}
\label{tab:hemt_bias}
\centering
\renewcommand{\arraystretch}{1.15}
\begin{tabular}{|l|c|c|}
\hline
Bias Points & Q1 (10\,mK stage) & Q2 (4\,K stage) \\
\hline
Drain–source voltage $V_{ds}$ & $45.8\,\text{mV}$ & $187\,\text{mV}$ \\
Drain–source current $I_{ds}$ & $50\,\mu\text{A}$ & $1.0\,\text{mA}$ \\
Gate voltage $V_{g}$ & $0\,\text{mV}$ & $-133\,\text{mV}$ \\
Dissipated power $P=V_{ds}I_{ds}$ & $2.3\,\mu\text{W}$ & $18.7\,\mu\text{W}$ \\
\hline

\end{tabular}
\end{table}

\section{Gain Estimation and Noise Characterization}

To measure the input noise of SPLENDOR’s cryogenic amplifier, we first characterized circuit components cryogenically~\cite{riley} and determined the amplifier gain using the MokuPro spectrum analyzer. The gain was extracted from frequency-response data by applying a sinusoidal input and fitting the measured transfer function to a simplified circuit model. This gain was then used, together with the known SR560 gain, to reference the measured output noise back to the amplifier input.

\subsection{Estimating Gain from Frequency Response}

\begin{figure*}[t]
\centering
\includegraphics[width=0.49\linewidth, trim=0cm 0.5cm 2.5cm 0cm, clip]{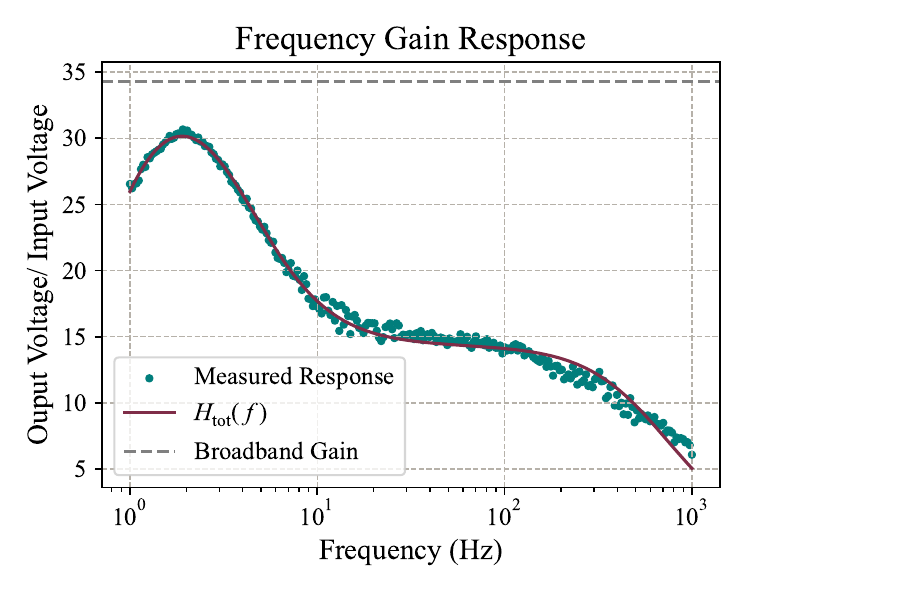}
\includegraphics[width=0.49\linewidth, trim=2cm 0cm 0cm 0cm, clip]{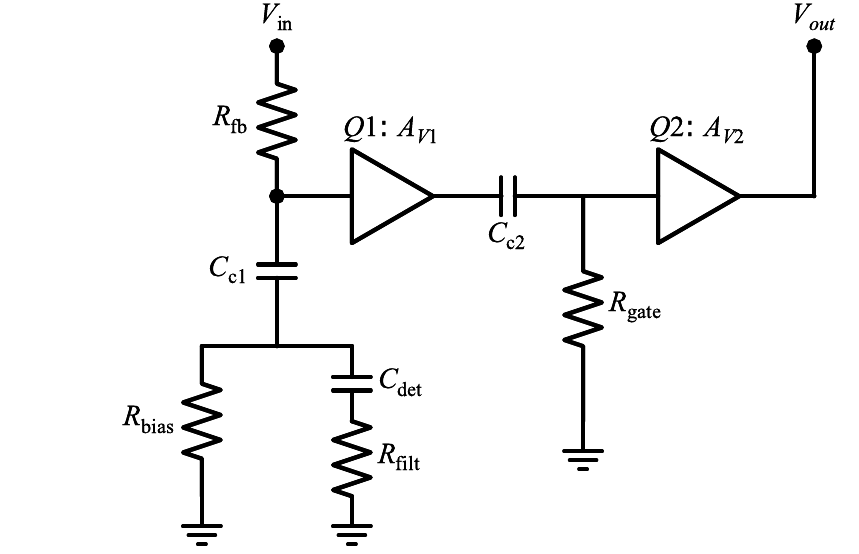}
\caption{ (Left) Frequency response data used to estimate broadband gain ($A_{V2}$) of the cryogenic amplifier shown in Figure \ref{fig: Schematic Offical}. Eq. \ref{eq:transferequation} is fit to the data with parameters shown in Table \ref{tab:gain fit}. (Right) Simplified circuit used to derive Eq. \ref{eq:transferequation} and estimated gain.  
}
\label{fig: freq response}
\end{figure*}

A series of AC signals (1\,Hz–1\,kHz) were injected at the gate of Q1 through the lines shown in Fig.~\ref{fig: Schematic Offical}. The measured frequency response (Fig.~\ref{fig: freq response}, left) was attenuated and shaped by the room-temperature injection board and 4\,K stage filtering, even though the true amplifier response is broadband and flat until rolling off near 100 kHz. To recover the intrinsic gain, the data were fit with the following transfer-function model: 

\begin{equation}
    \label{eq:transferequation}
    H_\text{total} \approx 
    A_{V1}\,A_{V2}\cdot \frac{j\omega C_\text{c2} R_\text{gate}}{1 + j\omega C_\text{c2} R_\text{gate}} \cdot \frac{Z_2(\omega)}{R_\text{fb}+Z_2(\omega)}
\end{equation}

\noindent This model (Eq.~\ref{eq:transferequation}) includes distinct terms corresponding to 
a voltage divider: 

\begin{equation}
    H_{VD}(\omega) = \frac{Z_2(\omega)}{Z_1 + Z_2(\omega)},
\end{equation}

\noindent where $Z_1 = R_\text{fb}$ and $Z_2(\omega)$ is the composite impedance formed by the bias resistor $R_\text{bias}$, coupling capacitor $C_\text{c1}$, filter resistor $R_\text{filt}$, and detector capacitance $C_\text{det}$. The full transfer function also includes a source-follower term, modeled as a unity-gain buffer:

\begin{equation}
    H_{SF} = A_{V1} \approx 1,
\end{equation}
as well as a term representing a high-pass network:

\begin{equation}
    H_{HP}(\omega) = \frac{j\omega C_\text{c2} R_\text{gate}}{1 + j\omega C_\text{c2} R_\text{gate}}.
\end{equation}
$H_{HP}(\omega)$  depends on the coupling capacitor $C_\text{c2}$ and gate resistance $R_\text{gate}$. Lastly, we also include a term corresponding to a HEMT operated in common-source configuration that acts as a voltage amplifier:

\begin{equation}
    H_{CS} = A_{V2}
\end{equation}
We note that the gain $A_{V2}$ \emph{is the parameter used to compute input-referred noise.} 
Constrained by measured cryogenic component values, the fit (shown in Fig.~\ref{fig: freq response}) returned a broadband gain $A_{V2}\approx34$ (Table~\ref{tab:gain fit}). This value was used in subsequent noise analysis.

\subsection{Input Referred Noise Model}

To model the input-referred noise, we restrict the noise sources to components at the base-temperature stage. Fig.~\ref{fig: noise} indicates that the amplifier is not limited by the Q2 input noise. At the input of Q1, the dominant voltage-noise contributions are:

(1) Johnson noise from the two 220~M$\Omega$ resistors in parallel,

\begin{equation}
    \label{eq: johnson noise}
    v_J^2(\omega) = 4 k_B T\,\operatorname{Re}\!\big[Z_\text{input}(\omega)\big],
\end{equation}
with

\begin{equation}
    Z_\text{input}(\omega) = \frac{R_\text{total}}{1 + j\omega R_\text{total} C_\text{total}}.
\end{equation}

(2) Amplifier pink (1/$f$) voltage noise,

\begin{equation}
    \label{eq: pink noise}
    v_\text{pink}^2(\omega) = 2\pi\frac{A^2}{\omega}.
\end{equation}

(3) Amplifier white voltage noise,

\begin{equation}
    v_\text{white}^2 = v_0^2.
\end{equation}

(4) Amplifier current noise with a frequency-dependent term and a white term,

\begin{equation}
    i_n^2(\omega) = \frac{B^2}{2\pi} \omega + i_0^2,
\end{equation}
which generates an equivalent input voltage noise of

\begin{equation}
    v_i^2(\omega) = \big|Z_\text{input}(\omega)\big|^2\, i_n^2(\omega).
\end{equation}

Combining these sources, the total input-referred voltage-noise spectral density is

\begin{multline}
    v_\text{total}^2(\omega) = 4 k_B T\,\operatorname{Re}\!\big[Z_\text{input}(\omega)\big] + 2\pi\frac{A^2}{\omega} \\ 
    + v_0^2 + \big|Z_\text{input}(\omega)\big|^2\!\left(\frac{B^2}{2\pi} \omega + i_0^2\right).
\end{multline}
The input referred voltage noise, and constituent noise model components, are shown in Fig.~\ref{fig: noise}. To convert the input-referred noise into an expected charge resolution, one must integrate over the effective signal bandwidth, which is set by the measured detector pulse dynamics (Sec.~\ref{sec:calibration}). We therefore defer a quantitative resolution estimate until after introducing the LED-based calibration.

\subsection{Temperature Dependence}

\begin{figure*}[t]
    \centering
    \includegraphics[width = \linewidth]{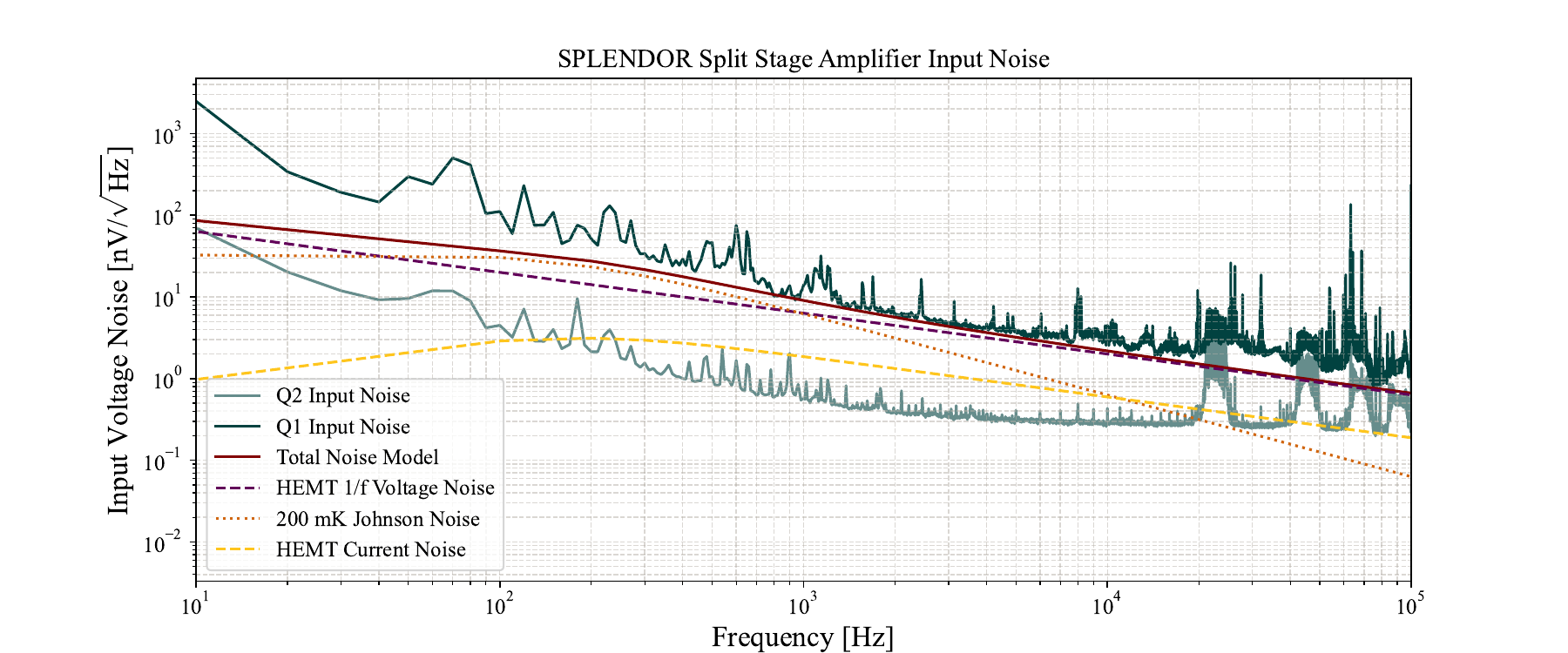}
    \includegraphics[width=0.84\linewidth, trim= 0cm 0cm 0cm 0cm, clip]{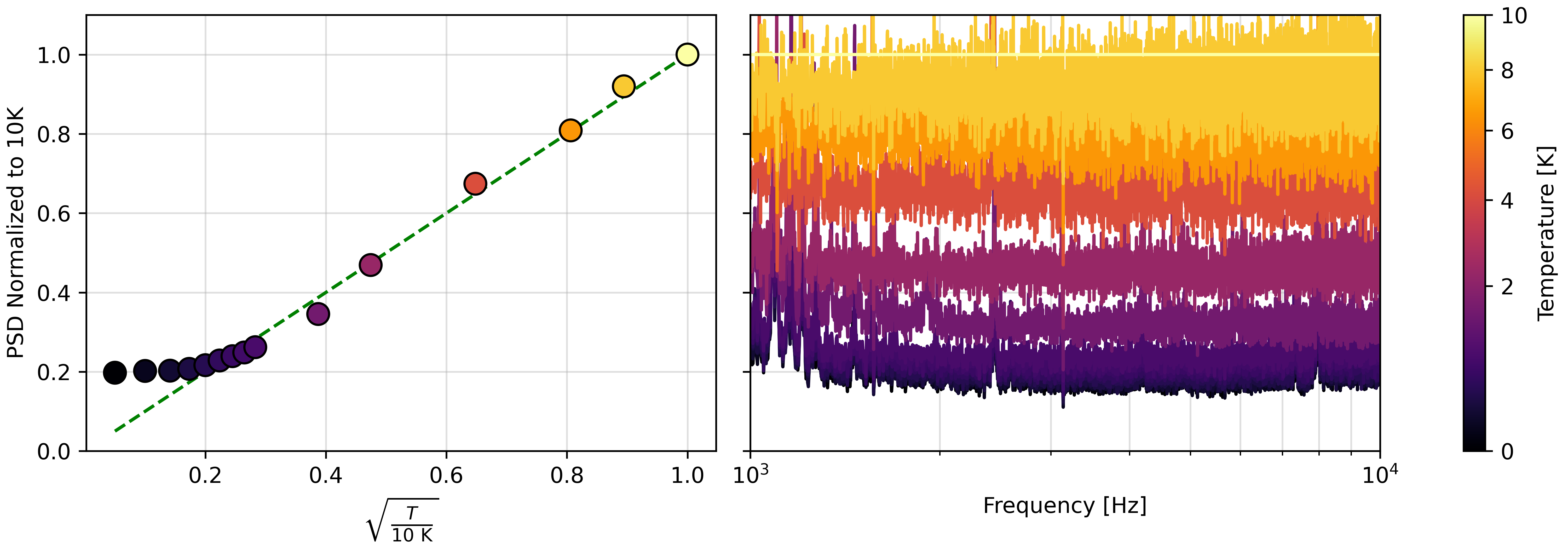}
    \caption{(Top) shows the input noise for Q1 and Q2 with each stage at 25\,mK and 4\,K respectively. The dominate noise sources are Johnson noise $v_\text{J}$ (eq. \ref{eq: johnson noise}) and 1/f voltage noise $v_\text{pink}$ (eq. \ref{eq: pink noise}) for Q1. Both components vary by $\sqrt{T}$. (Bottom Left) By normalizing the noise and plotting vs $\sqrt{T}$, you can see a divergence from the linear trend at 400\,mK. The green dashed line shows this expected linear trend. (Bottom Right) The flattest part of the spectrum, from 1\,kHz-10\,kHz, was normalized to the 10\,K noise. }
    \label{fig: noise}
\end{figure*}

As shown in Fig.~\ref{fig: noise} (top), the measured spectrum is dominated by Johnson noise and 1/f voltage noise from Q1. Fitting the model to the data required an effective noise temperature of 200\,mK, higher than the 25\,mK fridge base temperature, consistent with poor thermalization of the bias resistors.

A temperature sweep from 25\,mK to 10\,K showed that the Johnson-noise component increases with $\sqrt{T}$ until $\sim$400\,mK, where deviations appear (Fig.~\ref{fig: noise}, bottom). This further suggests that poor thermalization of the bias resistors and HEMT package are the cause of the elevated noise.

\begin{table}
\caption{Transfer Function Fit Parameters\label{tab:gain fit}}
\centering
\begin{tabular}{|c||c|c|}
\hline
Parameter & Value & Fit Status\\
\hline
$A_\text{V2}$ & $34.3\pm 0.3$ & Free\\
\hline
$A_\text{V1}$ & $1$ & Fixed\\
\hline
$R_\text{fb} $ & $285\pm 5~\text{M}\Omega$ & Free\\
\hline
$R_\text{bias} $ & $200~ \text{M}\Omega$ & Fixed\\
\hline
$R_\text{gate}$  & $100~\text{k}\Omega$& Fixed\\
\hline
$R_\text{filt} $ & $1 \text{k}\Omega$ & Fixed\\
\hline
$C_\text{c1} $ & $78 \pm 2~\text{pF}$ & Free\\
\hline
$C_\text{det} $ & $1.84\pm 3~\text{pF}$ & Free\\
\hline
$C_\text{c2}$ & $2.55\pm 9~\mu\text{F}$ & Free\\
\hline
\end{tabular}
\end{table}

\section{LED Shot Noise Calibration}
\label{sec:calibration}

\begin{figure*}
    \centering
    \includegraphics[height=0.311\textheight]{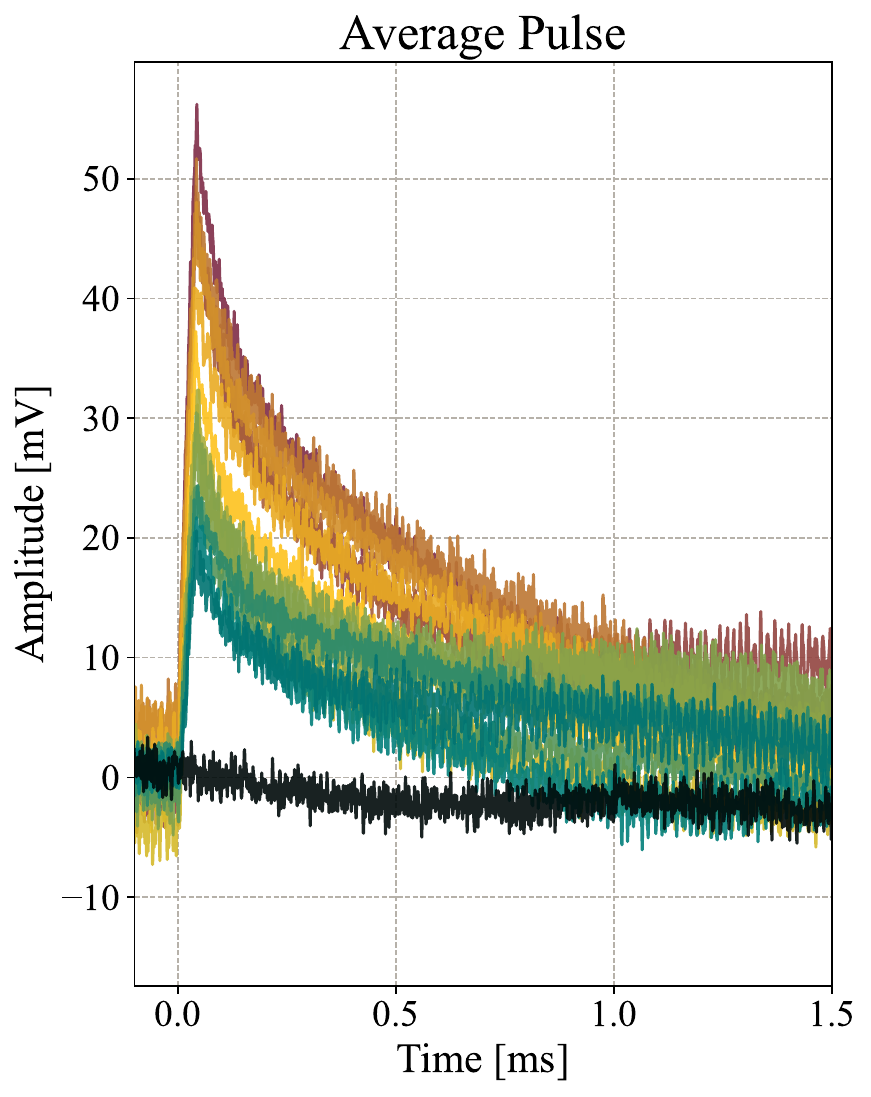}
    \includegraphics[height=0.311\textheight]{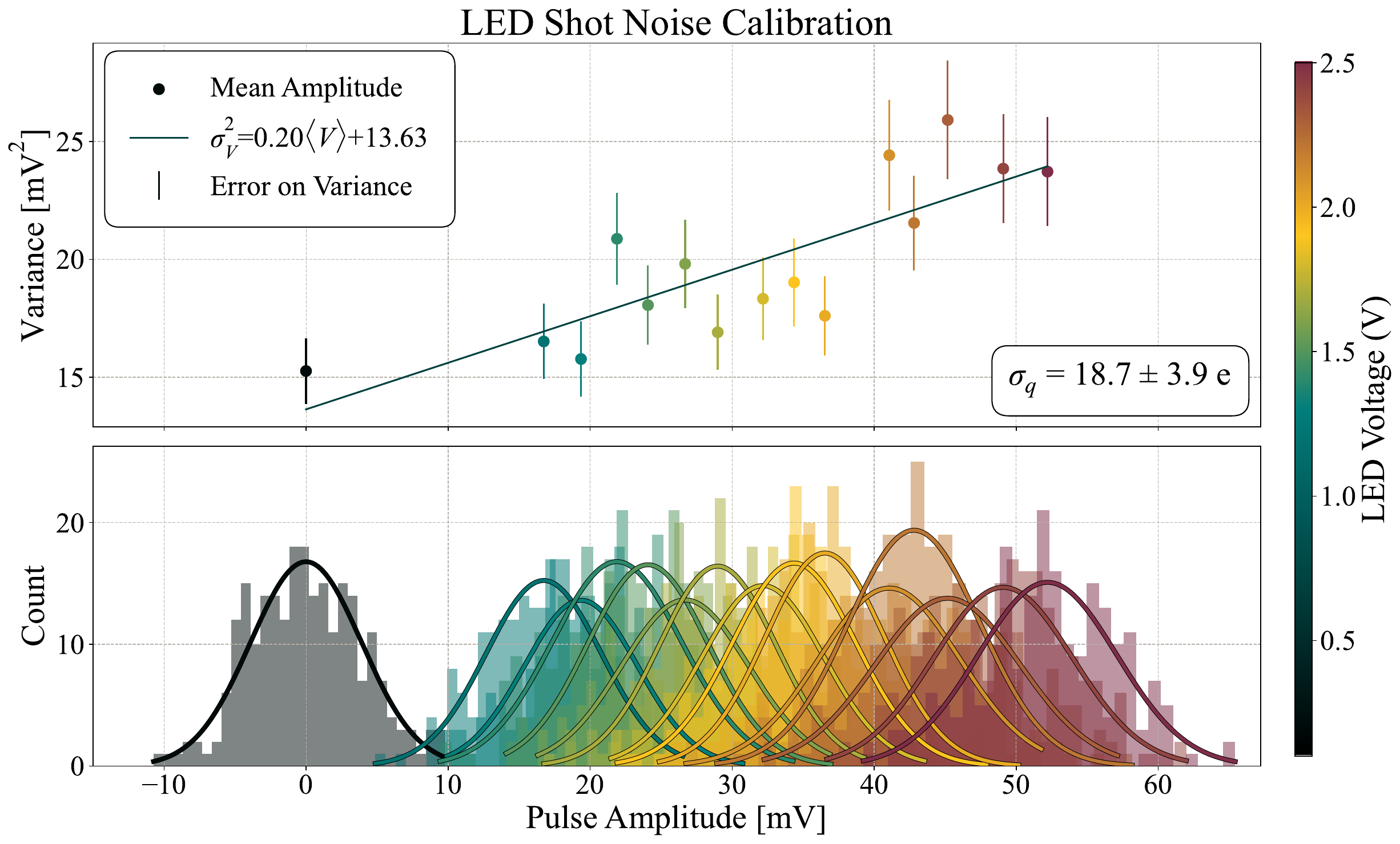}
    \caption{(Left) 300 filtered averaged pulses for 14 different LED intensities. (Bottom Right) Histogram of pulse heights for each LED intensity fitted to a gaussian. The color indicates the LED intensity. Black corresponds to zero point amplitude and variance. These data include 300 LED pulses of 625nm light, each with a 50\,$\mu$s pulse width. 5\,$\sigma$ quality cuts were done on both $\chi^2$ fits and pulse amplitudes. The pulse template used along with the optimal filter formalism was a 3 pole pulse with a one rise time and two fall times. (Top Right) Linear relationship between mean pulse amplitude and variance based on equation \ref{eq:linfit}. From fit parameters and equation \ref{eq: charge resolution}, a $19\,\pm 4 e^-$ resolution is measured. }
    \label{fig: LED calibration}
\end{figure*}

\subsection{Shot Noise Model}
To establish the absolute charge sensitivity of the amplifier, we used Poissonian shot noise from a 625\,nm Thorlabs M625F2 fiber-coupled LED as a calibration source. The LED was driven at room temperature by a MokuPro output (Fig.~\ref{fig: Schematic Offical}) and coupled through an Accuglass fiber feedthrough. Between temperature stages of the dilution refrigerator, multimode cryogenic optical fibers were SMA-coupled for thermal anchoring, introducing stage-by-stage attenuation. At base temperature, the fiber illuminated the entire detector cavity diffusely, ensuring uniform exposure of the detector substrates. For calibration, a silicon point-contact target was chosen due to its well-understood photon interactions~\cite{Ramanathan:2020fwm}.

The LED was pulsed at 30 Hz for 10 seconds at 14 driver power settings, producing distinct distributions of detected charge for each power. The amplifier output is related to the number of absorbed photons $n_\gamma$ by

\begin{equation}
V_\text{out} = \frac{G}{C_\text{in}} \varepsilon \left[n_\gamma \eta + \delta_q\right],
\end{equation}
where $G$ is the total gain of the amplifier chain, $C_\text{in}$ is the input capacitance, $\eta$ is the photon-to-charge conversion (unity for 625 nm photons), $\varepsilon$ is the charge collection efficiency, and $\delta_q$ represents intrinsic detector charge noise.

Because $n_\gamma$ follows Poisson statistics, the mean and variance of $V_\text{out}$ scale as

\begin{align}
\langle V_\text{out}\rangle &= \frac{G}{C_\text{in}}\eta\varepsilon\langle n_\gamma\rangle, \
\sigma^2_{V\text{out}} &= \left(\frac{G}{C_\text{in}}\varepsilon\right)^2\left[\eta^2 \langle n_\gamma\rangle + \sigma_q^2\right].
\end{align}
Rearranging gives a simple linear relation between variance and mean output voltage,

\begin{equation}
\sigma^2_{V\text{out}} = m \langle V_\text{out}\rangle + b,
\label{eq:linfit}
\end{equation}
with slope and intercept defined as

\begin{equation}
m = \frac{G}{C_\text{in}}\eta\varepsilon, \quad
b = \left(\frac{G}{C_\text{in}}\varepsilon\right)^2\sigma_q^2.
\end{equation}
Since each 625\,nm photon generates a single electron-hole pair ($\eta = 1$), the intrinsic charge noise is obtained directly from the fit parameters:

\begin{equation}
\sigma_q = \sqrt{\frac{b}{m^2}}.
\label{eq: charge resolution}
\end{equation}
This relation provides the baseline charge resolution of the amplifier, independent of the photon flux calibration.

\subsection{Data Collection and Processing}

For each LED power setting, the detector was illuminated with $50\,\mu$s pulses of 625\,nm light at 30\,Hz for 10\,s, yielding 300 pulses per setting. A 9\,V bias was applied across the silicon detector, corresponding to an 18\,V/mm electric field.

Pulse amplitudes were extracted using the optimal filter formalism~\cite{RADEKA196786}\cite{Kurinsky_2018}, implemented in the \texttt{splendsp} and \texttt{splendaq} packages~\cite{watkins2023splendaq}. The optimal filter requires three inputs: the pulse time, baseline noise, and a pulse template. The LED trigger times were recorded in the timestream, and baseline noise was measured between LED pulses. Pulse templates were constructed by averaging all 300 pulses at a given LED intensity and fitting to a three-pole function (one rise time, two fall times). For intensities with visible pulses, the templates were consistent, allowing a single template to be used across all LED settings. The fitted fall time corresponded to an input capacitance of 8\,pF, set by the parallel combination of the bias and feedback resistors with the total amplifier input capacitance.

\begin{table}[h]
\caption{Input Noise Fit Parameters}
\label{tab:table2}
\centering
\begin{tabular}{|c|c|c|c|c|c|c|}
\hline
$T$ & $R_\text{input}$ & $C_\text{input}$ & $A$ & $v_0$ & $B$ & $i_0$ \\
\hline
$200\,\text{mK}$ & $110\,\text{M}\Omega$ & $8\,\text{pF}$ & $200\,\text{nV}$ & $0\,\text{nV}$ & $3\,\text{aA}$ & $0\,\text{aA}$ \\
\hline
\end{tabular}
\end{table}

For each event, the optimal filter returned an amplitude and $\chi^2$ value. Outliers more than $5\sigma$ from the mean in either $\chi^2$ or amplitude were rejected. From the cleaned distributions, the mean $\langle V_\text{out}\rangle$ and variance $\sigma_{V_\text{out}}^2$ were computed for each LED power setting. These results are shown in Fig.~\ref{fig: LED calibration}, along with a linear fit to Eq.~\ref{eq:linfit}, yielding an intrinsic charge resolution of $19 \pm 4\,e^-$.

\subsection{Comparison of Noise-Derived and Calibrated Resolution}

Using the measured fall time of the LED pulses, the effective bandwidth is $f_\mathrm{BW} \approx 1/(2\pi \tau)$. Integrating the modeled input-noise spectrum over this bandwidth (as done in Ref.~\cite{anczarski2023twostage} and \cite{Phipps2016}) yields an expected charge resolution of $\sim 17\,e^-$, consistent with the $19 \pm 4\,e^-$ obtained from LED calibration. This agreement confirms that the amplifier is thermally limited and that the noise model captures the dominant contributions.

\section{Discussion and Outlook}

We demonstrated that our measured amplifier performance is thermally limited and achieves a consistent charge resolution of $\sim 20\,e^-$. The resolution inferred from input-referred noise agrees with that from LED shot-noise calibration, confirming full charge collection. With this resolution, materials with band gaps below 50 meV become accessible, enabling charge-mediated probes of sub-eV depositions beyond the $\sim 1$ eV threshold of conventional semiconductors.  

Thermalization improvements, such as optimized bias resistor contacts or alternative active-reset strategies \cite{PHIPPS2019181}, could plausibly reduce the resolution toward 3\,$e^-$, approaching the single-electron regime. In parallel, a complete room temperature electronics board is funded for development which can eliminate the SR560 and improve noise performance. Even without such upgrades, the present amplifier provides an immediately deployable platform for characterizing new narrow-gap semiconductors, conducting exploratory sub-MeV dark matter searches, and surveying condensed-matter materials at sub-Kelvin temperatures.  

It also may be possible that the current noise for these low-capacitance HEMTs deviates strongly from the higher capacitance trends in this operating regime. We note that the noise in the low-intermediate frequency ranges is limited by a combination of 60~Hz, low-frequency vibration, and Johnson noise, and that the high-frequency regime we use to constrain the pink noise could equally well be fit by a substantially elevated HEMT current noise. Future measurements in the absence of the gate impedance will be used to shine futher light on whether this is an additional noise component, or if these HEMTs do suffer from abnormally high current noise due to low bias power or another effect.

Looking ahead, reaching true single-electron sensitivity will likely require alternative device concepts based on superconducting quantum devices. These approaches are complementary to the HEMT amplifier described here and represent a longer-term pathway toward quantum-limited charge readout. In the meantime, the demonstrated HEMT-based design fills an important role: enabling low-noise, modular charge readout for rare-event searches and quantum materials studies.

\section*{Acknowledgments}
The authors are grateful to Jim MacArthur for the design and production of the HEMT Power Supply (HPS) used in this setup. This work was supported by the Laboratory Directed Research and Development program of Los Alamos National Laboratory under project numbers 20220135DR, 20220252ER, 20230777PRD1, and 20230782PRD1. Los Alamos National Laboratory is operated by Triad National Security, LLC, for the National Nuclear Security Administration of the U.S. Department of Energy (DOE) under contract number 89233218CNA000001. NAK was supported in part by the DOE Early Career Research Program (ECRP) under FWP 100872. JA was supported in part by a Kavli Institute for Particle Astrophysics and Cosmology Chabollah Fellowship. AP and OA were supported in part by the U.S. Department of Energy under award DE-SC0025747.



\end{document}